# Light-induced switching in pDTE – FICO 1D photonic structures

Ilka Kriegel [1], Francesco Scotognella [2,3,*]

[1]*Department of Nanochemistry, Istituto Italiano di Tecnologia (IIT), via Morego, 30, 16163 Genova, Italy*
[2]*Dipartimento di Fisica, Politecnico di Milano, Piazza Leonardo da Vinci 32, 20133 Milano, Italy*
[3]*Center for Nano Science and Technology@PoliMi, Istituto Italiano di Tecnologia, Via Giovanni Pascoli, 70/3, 20133, Milan, Italy*
*email address: francesco.scotognella@polimi.it

**Abstract**
We propose the design of 1D photonic crystals and microcavities in which fluorine-indium codoped cadmium oxide (FICO) nanocrystal based layers and layers of diarylethene-based polyester (pDTE) are alternated or embedded in a microcavity. The irradiation with UV light results in two different behaviours: i) it dopes the FICO nanocrystals inducing a blue shift of their plasmonic resonances; ii) it changes the real part of the refractive index of the photochromic pDTE polymer. These two behaviours are combined in the proposed photonic structures and can be useful for switchable filters and cavities for light emission.

**Introduction**
The tuning of the band gap in photonic structures, in which such gap arises due to a modulation of the refractive index [1–3], can be of great importance for light emitting devices and systems in which a colour control is desirable. Recent review articles report the possibility to tune the optical response with a highly effective external stimulus, such as the electric field [4–8]. Some of the latest articles describe reflectors based on chiral liquid crystals in polymer network that respond to electric and thermal stimuli [9] and electroresponsive photonic structures that combine structural and electrochromic effects [10].
Other materials employed in photonic structures to enable tuning are photochromic materials [11,12], or heavily-doped semiconductor nanocrystals [13,14]. The optical properties of the latter can be manipulated via photodoping by exciting them with light in the region of their bandgap. Examples are fluorine-indium codoped cadmium oxide (FICO) [15], $In_2O_3$ and Sn-doped $In_2O_3$ nanocrystals [16], $Fe^{3+}$-doped ZnO nanocrystals [17].
The use of light instead of electricity can pave the way to the fabrication of contactless switches. In this paper, we propose the design of a 1D photonic crystal made by alternating two functional layers. The first layer is a $SiO_2$ matrix with FICO nanoparticles embedded, while the second layer is a photochromic polymer. The two materials act differently upon UV irradiation: FICO nanoparticles are photodoped by UV light, with a subsequent increase of carrier density that results in a blue shift of their near infrared plasmon resonance. The photochromic polymer, a diarylethene-based polyester (pDTE) in this case [18], shows an increase of the refractive index. Thus, a photonic crystal containing pDTE shows a red shift of the photonic band gap upon UV irradiation. Taking advantage of both effects in a photonic crystal structure composed of FICO nanoparticles and pDTE a simultaneous red shift of the photonic band gap together with a blue shift of the plasmon resonance of FICO can be induced. The proposed microcavities in our work, in which a defect composed of $FICO:SiO_2$ and pDTE is embedded between two photonic crystals of $SiO_2$ and $TiO_2$ are able to show either a tuning of the defect mode in the cavity or the cross-over of different transmission peaks.

**Methods**

We have extracted, from Toccafondi et al. [18], the Sellmeier equation of p-DTE in the transparent phase:

$$n^2(\lambda) - 1 = \frac{1.194\lambda^2}{\lambda^2 - 0.1175} + \frac{0.06976\lambda^2}{\lambda^2 - 0.5233} + \frac{0.09367\lambda^2}{\lambda^2 - 0.2076} \tag{1}$$

and its Sellmeier equation of p-DTE in the blue phase:

$$n^2(\lambda) - 1 = \frac{0.02015\lambda^2}{\lambda^2 - 0.7242} + \frac{1.209\lambda^2}{\lambda^2 - 0.01433} + \frac{0.2905\lambda^2}{\lambda^2 - 0.4754} \tag{2}$$

where λ is in micrometers. This corresponds to a refractive index, at 1 μm, of 1.545 for the transparent phase and 1.606 for the blue phase of p-DTE, respectively. An important parameter that should be considered in pDTE is the variation of the polymer layer thickness from the transparent phase to the blue phase. In Ref. [18] it is reported a decrease of film thickness, due to film density change, of about 1.5 % from the transparent phase to the blue phase.

The parameters related to the optical properties of FICO nanoparticles are taken from Kriegel et al. [15]. We consider a layer of FICO nanoparticles embedded in a $SiO_2$ matrix (for the $SiO_2$ we consider, in the studied range, a dielectric constant of 2.1025).

To predict the behaviour of the plasmonic response in our photonic structures the Drude model can be employed [19], where the frequency dependent complex dielectric function of FICO can be written as:

$$\varepsilon_{FICO}(\omega) = \varepsilon_1(\omega) + i\varepsilon_2(\omega) \tag{3}$$

where

$$\varepsilon_1 = \varepsilon_\infty - \frac{\omega_P^2}{(\omega^2 - \Gamma^2)} \tag{4}$$

and

$$\varepsilon_2 = \frac{\omega_P^2 \Gamma}{\omega(\omega^2 - \Gamma^2)} \tag{5}$$

where $\omega_P$ is the plasma frequency and $\Gamma$ represents the free carrier damping [20]. For FICO nanoparticles we used $\omega_P = 2.3196 \, eV$ and $\Gamma = 0.07 \, eV$ [13,15].

The dielectric function of the FICO nanoparticle film (a network of necked nanoparticles with $SiO_2$-filled pores) can be described by the Maxwell-Garnett effective medium approximation [21–23], which is given by

$$\varepsilon_{eff,FICO} = \varepsilon_{Air} \frac{2(1-\delta_{FICO})\varepsilon_{SiO_2} + (1+2\delta_{FICO})\varepsilon_{FICO}}{2(2+\delta_{FICO})\varepsilon_{SiO_2} + (1-\delta_{FICO})\varepsilon_{FICO}} \tag{6}$$

where $\delta_{FICO}$ accounts for the volume fraction occupied by the FICO nanoparticles. The refractive index of the FICO:$SiO_2$ layer is then given by: $n_{eff,FICO}^2 = \varepsilon_{eff,FICO}$ . The electric field distribution inside the sample was determined by employing the method proposed in Ref. [24].

**Results and Discussion**
Experimental results have shown that FICO nanoparticles are photodoped when excited with UV light (at 400 nm with a 150 fs pulsed laser) leading to an increased carrier density of up to around 7% and the concomitant blue shift of their near infrared plasmon resonance [15]. Photodoping is related to an increase of the conduction band carrier density when exciting the material above the bandgap energy. This additional carrier density adds to the charge density of the n-type material, i.e. FICO nanoparticles. Instead, the pDTE polymer shows a transition from a transparent phase to a blue phase, with a variation of the real part of the refractive index $\Delta Re(n)$ of 0.05 in the infrared region when impinged with a UV lamp at 311 nm [18]. In this work we take advantage of these effects, by combining both materials into a photonic structure. In the photonic crystal, of 20 bilayers, the thickness of the $FICO:SiO_2$ layer is 145 nm, while the thickness of pDTE layer is 355 nm in the transparent phase and 350 nm in the blue phase (due to the film density change induced by the transition from the open form to the closed form of pDTE [18]). The proposed structures can be experimentally fabricated by employing sputtering for $SiO_2$-$TiO_2$ reflectors [25], and spin coating for $FICO:SiO_2$ and pDTE.

In Figure 1 the black curve (u/t) displays the transmission spectrum of the photonic structure for undoped FICO nanoparticles and the transparent phase of pDTE, while the red dashed curve (d/b) displays the spectrum for the case of doped FICO nanoparticles and blue phase of pDTE, i.e. after impinging the device structure with UV light. In both structures a dip in the transmission around 0.8 eV is observed due to the photonic bandgap. When switching is induced due to excitation with UV light, the photonic bandgap is shifted to higher energies.

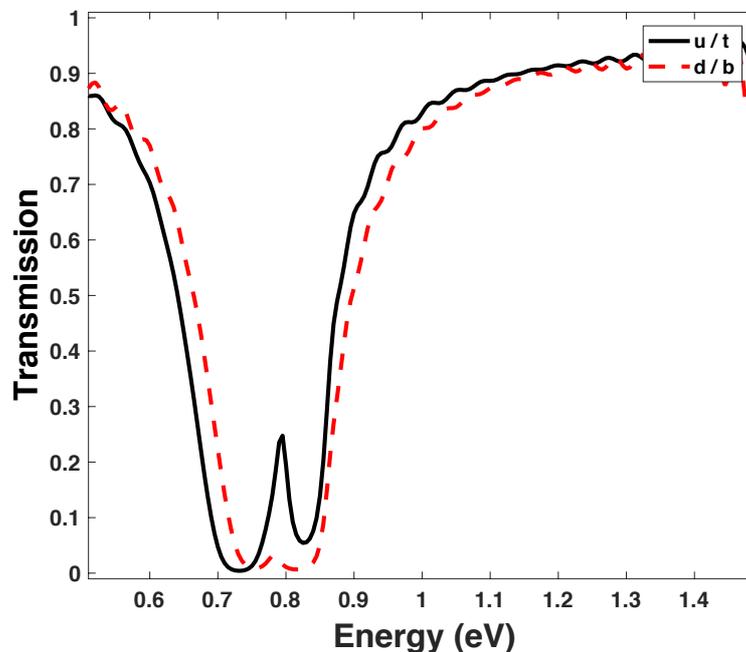

**Figure 1.** Transmission spectra of the $FICO:SiO_2$/pDTE photonic crystals. The photonic crystal with undoped FICO nanoparticles and the transparent phase of pDTE (u/t) is described by the black solid curve, while the one with doped FICO nanoparticles and blue phase of pDTE (d/b) is described by the red dashed curve.

Notably, the undoped/transparent photonic crystal shows a transmission peak at about 0.8 eV that is removed when the photonic crystal is transferred to the doped/blue condition. This can be understood by the change in refractive index contrast between the alternating layers that shifts towards the red the photonic band gap, together with a blue shift of the plasmonic resonance of FICO because of the photo-induced increase of the carrier density.

Thus, an optical switch is created triggered entirely by light. We remark here that the photonic bandgap and the transmission dip can be designed largely, by controlling the thickness of the respective layers, giving the option plan a device structure in the wavelength range of interest.

We would like to highlight the importance to have an optimal surface roughness of the layers to fabricate photonic crystals with consistent and reproducible light transmission spectra. To simulate the surface roughness of the layers, we generate a random thickness of the layers with values that follow a Gaussian function

$$f(x) = e^{\frac{-(x-c)^2}{2\sigma^2}} \qquad (7)$$

in which $x, c, \sigma$ are in nanometers. $c$ is 145 nm for FICO:SiO$_2$ and 355 nm for p-DTE, respectively. $\sigma$ is related to the surface roughness of the layers. For three values of $\sigma$ (5, 10 and 20), we calculated the transmission spectra of 10 FICO:SiO$_2$ photonic crystals with a different sequence of layer thicknesses. In Figure 2 we show the transmission spectra, highlighting the fluctuation of the spectral position of the photonic band gap for $\sigma = 10$ and $\sigma = 20$, especially in the case of undoped FICO / transparent p-DTE (u/t) photonic crystals. In literature, spin coated p-DTE films have surface roughness below 1 nm [18], while SiO$_2$ spin coated films have surface roughness below 4 nm [26], ensuring the fabrication of high optical quality photonic crystals.

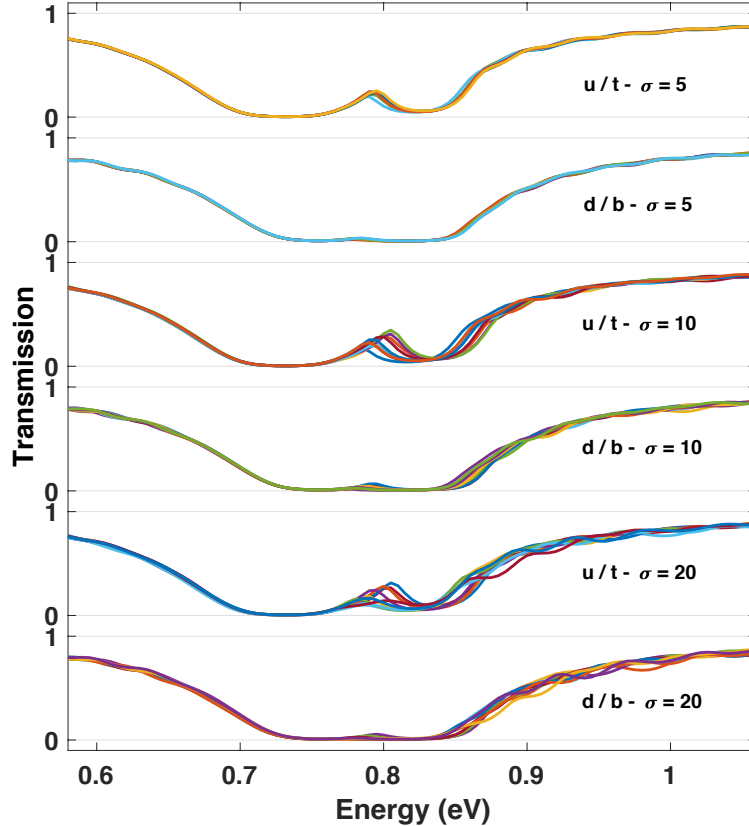

**Figure 2.** Transmission spectra of the u/t and d/b FICO:SiO$_2$/pDTE photonic crystals for different values of $\sigma$.

We further investigated a photonic microcavity (Figure 3), in which the defect is composed by a layer of FICO:SiO$_2$ and a layer of pDTE, where the thickness of the FICO:SiO$_2$ layer is 180 nm, while the thickness of pDTE layer is 180 nm in the transparent phase and 177 nm in the blue phase. The defect is sandwiched between two SiO$_2$-TiO$_2$ photonic crystals of 5 bilayers each (in this order: [SiO$_2$-TiO$_2$]$_5$-FICO:SiO$_2$-pDTE-[TiO$_2$-SiO$_2$]$_5$ ). The thickness of SiO$_2$ is 219 nm, while the one of TiO$_2$ is 150 nm. The wavelength dependent refractive indexes were taken from Ref. [27] for SiO$_2$ and from Ref. [28] for TiO$_2$. In Figure 3 is given the resulting transmission spectrum for the device structure in both, the original (black curve, u/t) and the switched state (red curve, d/b), where the sharp transmission due to the cavity gap is dominating the spectrum. In Figure 3b a zoom into this spectral region is shown, demonstrating the shift of this sharp resonance by several nanometers to the blue, induced entirely by excitation with light.

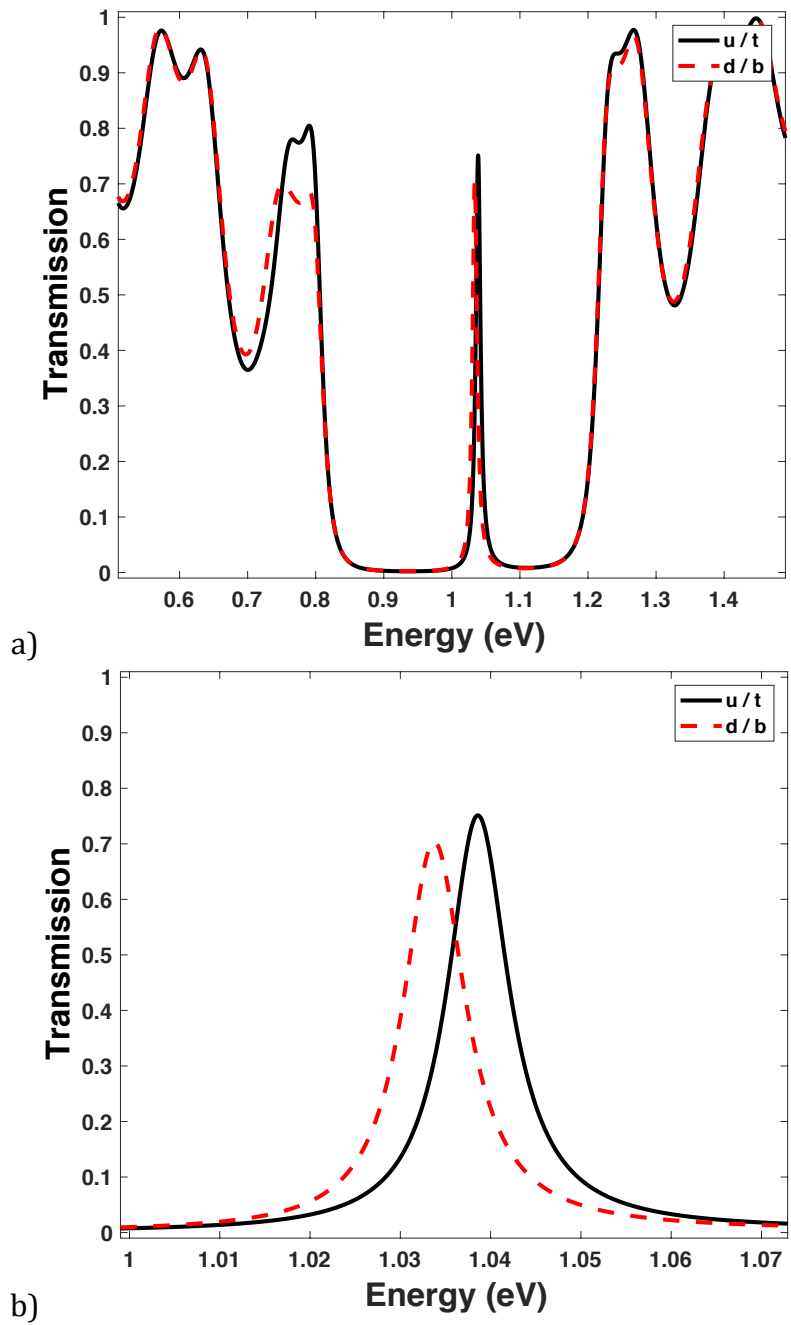

a)

b)

**Figure 3.** (a) Transmission spectra of the [SiO$_2$-TiO$_2$]$_5$-FICO:SiO$_2$-pDTE-[TiO$_2$-SiO$_2$]$_5$ microcavity in the undoped/transparent (u/t) and in the doped/blue (d/b) conditions. (b) Zoom of the transmission spectra in the cavity defect region.

A closer look into the electric field distribution within the photonic structure (Figure 4) reveals the enhancement of the field, in the spatial region corresponding to the defect of the cavity, being more pronounced for the undoped/transparent condition, while UV excitation results in a decrease of the field.

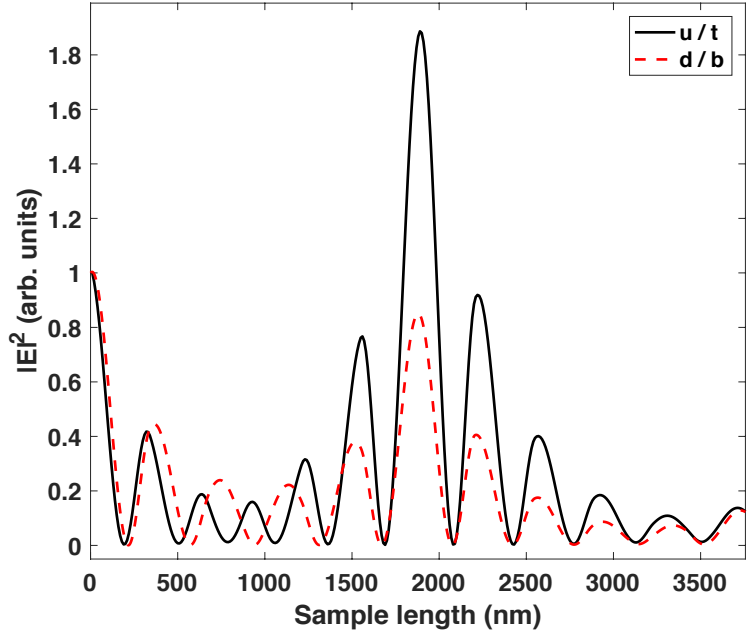

**Figure 4.** Electric field distribution (normalized by the value at the beginning of the sample) of the [SiO$_2$-TiO$_2$]$_5$-FICO:SiO$_2$-pDTE-[TiO$_2$-SiO$_2$]$_5$ microcavity.

The great advantage of our structure is the flexibility in device design. Therefore we studied a second photonic microcavity (Figure 5), in which the defect is composed by a thicker layer of FICO:SiO$_2$ and a thinner layer of pDTE, i.e. the FICO:SiO$_2$ layer is 1100 nm, while the pDTE layer is 400 nm (in the transparent phase and 394 nm in the blue phase). The defect is between two SiO$_2$-TiO$_2$ photonic crystals of 10 bilayers each (in this order: [SiO$_2$-TiO$_2$]$_{10}$-FICO:SiO$_2$-pDTE-[TiO$_2$-SiO$_2$]$_{10}$ ). The thickness of SiO$_2$ is 219 nm, while the one of TiO$_2$ is 150 nm.

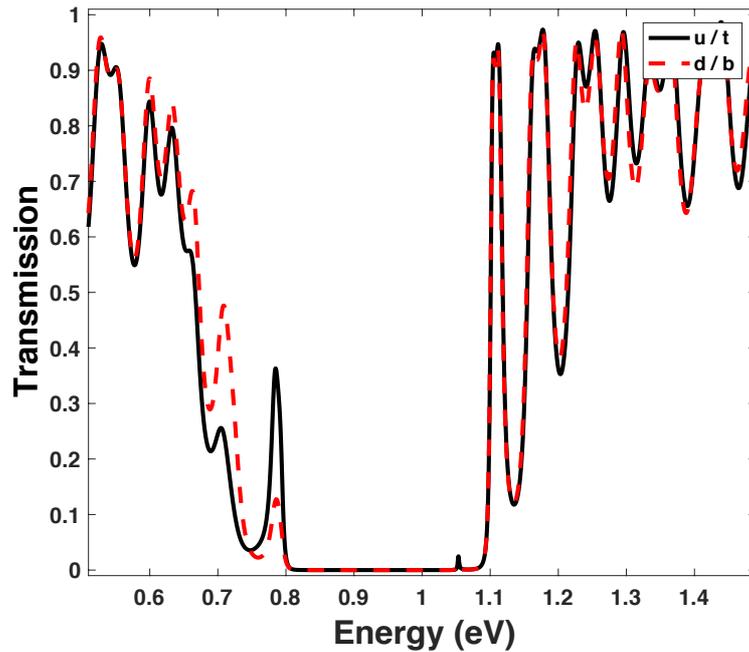

**Figure 5.** Transmission spectra of the $[SiO_2\text{-}TiO_2]_5$-$FICO\text{:}SiO_2$-pDTE-$[TiO_2\text{-}SiO_2]_5$ microcavity in the undoped/transparent (u/t) and in the doped/blue (d/b) conditions for a device with a thicker $FICO\text{:}SiO_2$ and a thinner pDTE film within the defect layer.

In this structure the transmission spectrum is dominated by a pronounced microcavity around 0.78 eV (black curve in Figure 5). With UV irradiation we can observe in this second microcavity an increase of the transmission intensity for the peak at 0.71 eV and a decrease for the peak at 0.78 eV, inducing a cross-over in transmission peaks.

**Conclusion**
In this paper, we have proposed the design of a 1D photonic crystal switchable by optical triggers only. For this we employ two different active layers that show an optical response to UV irradiation: The first one is a FICO nanoparticle doped $SiO_2$ matrix, while the second layer is made by a photoswitchable polymer film (pDTE). The FICO nanocrystals can be optically manipulated via photodoping with UV light, with a subsequent increase of carrier density and a blue shift of the near infrared plasmon resonance. The pDTE film instead shows an increase of the real part of the refractive index upon UV light exposure. We proposed two different types of structures, i.e. a photonic crystal with alternating layers and microcavities in which a defect made with $FICO\text{:}SiO_2$ and pDTE is embedded between two photonic crystals of $SiO_2$ and $TiO_2$ . We have shown that the light transmission through these photonic structures can be controlled and actively manipulated by UV light with the flexibility of device design. We additionally highlight the option of ultrafast photodoping in FICO nanoparticles, as presented in Ref. [15] that within a similar structure as proposed by us can be of interest for ultrafast all-optical signal processing.

**Acknowledgements**
This project has received funding from the European Union's Horizon 2020 research and innovation programme (MOPTOPus) under the Marie Skłodowska-Curie grant agreement No. [705444], as well as (SONAR) grant agreement no. [734690].